\documentclass[12pt]{iopart}
\usepackage{amssymb}
\usepackage{cite}
\usepackage{graphicx}
\usepackage{color}

\newcommand {\E}  {{\varepsilon}}
\newcommand {\om}  {{\omega}}
\newcommand {\Om}  {{\Omega}}

\newcommand {\dUmax} {U^{\prime}_{\max}}

\newcommand {\Ld} {{L_{\rm d}}}
\newcommand {\La} {{L_{\rm a}}}
\newcommand {\Lrad} {{L_{\rm r}}}

\newcommand {\Nd} {{N_{\rm d}}}

\begin{document}

\title[Electron-based crystalline undulator]
{An electron-based crystalline undulator}

\author{
M Tabrizi$^{1}$, 
A V Korol$^{1,2}$, 
A V Solov'yov$^{1}$ 
and 
Walter Greiner$^{1}$
}

\address{
$^1$ Frankfurt Institute for Advanced Studies, Johann Wolfgang 
G\"othe-Universit\"at,
 Max-von-Laue-Str. 1, 60438 Frankfurt am Main, Germany}
\address{
$^2$ Department of Physics, St. Petersburg State Maritime 
Technical University, Leninskii prospect 101,
St. Petersburg 198262, Russia}
\ead{tabrizi@bk.ru, a.korol@fias.uni-frankfurt.de, 
solovyov@fias.uni-frankfurt.de}

\begin{abstract}
We discuss the features of a crystalline undulator of the novel type  
based  on the effect of a planar channeling of ultra-relativistic 
{\em electrons}  in a periodically bent crystals.
It is demonstrated that an electron-based undulator is feasible in the 
tens of GeV range of the beam energies, which is noticeably higher  
than the energy interval allowed in a positron-based undulator.
Numerical analysis of the main parameters of the undulator as well as the 
characteristics of the emitted undulator radiation is carried out 
for 20 and 50 GeV electrons channeling in diamond and silicon crystals 
along the (111) crystallographic planes.  
\end{abstract}
\maketitle

\section{Introduction \label{Introduction}}
In this paper we suggest and discuss a new type of a powerful source
of high energy photons generated by a bunch of ultra-relativistic
electrons undergoing planar channeling in a periodically bent
crystal. 
We call such system 'an electron-based crystalline' undulator. 
The feasibility of the undulator has been recently proven
for the first time in \cite{TKSG2006}. In this work the results of more
detailed qualitative and quantitative analysis of the radiation
formed in electron-based crystalline undulators at different
energies of electron and for different types of crystals are
presented.

As it is known \cite{KSG1998,KSG1999}, in a crystalline undulator there
appears, in addition to the well-known channeling radiation
\cite{Kumakhov2}, the radiation of an undulator type which is due to the
periodic motion of channeling particles which follow the bending of
 the crystallographic planes. 
In the cited papers, as well as in the subsequent publications 
\cite{KKSG2000_tot,KSG2000_loss,KSG2001_Dech,KSG2004_review}, 
a feasibility to create a short-wave 
crystalline undulator emitting intensive and monochromatic radiation, 
based on an ultra-relativistic positron channeling was proven 
(for the latest review  see \cite{KSG2004_review}). 
More recently\cite{KSG2005_SPIE1} it was shown that the 
brilliance of the radiation from a positron-based crystalline
undulator in the energy range from  hundreds of keV up to tens
of MeV is comparable to that of a conventional light source
of the third generation but for much lower photon energies.
Experimental study of this phenomenon is on the way within the framework 
of the PECU project 
\cite{PECU}.

\begin{figure}[ht]
\centering
\includegraphics[clip,scale=0.6,angle=0]{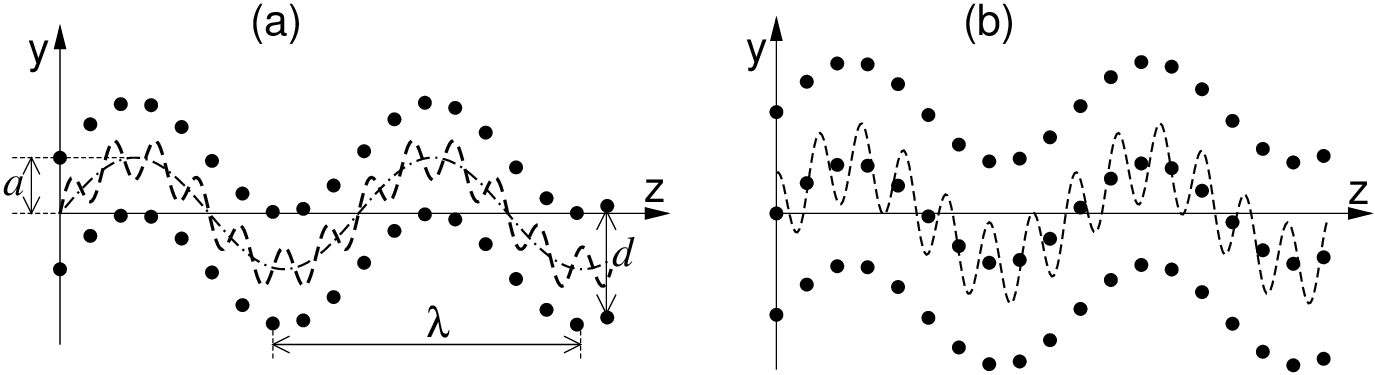}
\caption
{Schematic representation of the positron-based 
(graph (a)) and the electron-based (graph (b)) crystalline
undulators. Circles denote the atoms belonging to neighbouring
crystallographic planes (separated by the distance $d$) which are
periodically bent. The wavy lines show the trajectories of
channeling particles. A positron channels between two neighbouring
planes, whereas the electron channeling occurs nearby the
crystallographic plane. The profile of periodic bending is given by
$y(z)= a\sin (2 \pi z /\lambda)$, where $\lambda$ and $a$ are the
period and amplitude of the bending (indicated in graph (a)).
} 
\label{figure.1}
\end{figure}

The operational principle of a crystalline undulator does not depend
on the type of a projectile and is illustrated by figure
\ref{figure.1}. Under certain conditions \cite{KSG1998,KSG1999} an
ultra-relativistic charged particle, which enters the crystal at
the angle smaller than the Lindhard's critical angle \cite{Lindhard}, 
will penetrate through the crystal following the bendings of its planes.
Consequently, the trajectory of the particle contains two elements.
Firstly, there are oscillations due to the action of the interplanar
force, - the channeling oscillations\cite{Lindhard}, whose frequency
$\Om_{\rm ch}=c\sqrt{2\dUmax/d \E}$ 
($c$ is the speed of light) depends on the energy of the particle $\E$ and
the parameters of the channel: the maximal gradient of the
interplanar potential $\dUmax$  and the interplanar distance $d$. 
Secondly, there are oscillations due to the periodicity of the
bendings, the undulator oscillations, whose frequency is
$\om_{0}\approx 2\pi c/\lambda$, where $\lambda$ is the spatial
period of bending. 
The spontaneous emission is associated with both of these oscillations. 
Typical frequencies of the channeling and undulator radiation are 
$\om_{\rm ch}\approx2\gamma^{2} \Om_{\rm ch}$ and
$\om\approx4\gamma^{2}\om_{0}/(2+p^{2})$, where
$\gamma=\varepsilon/mc^2$ is the Lorentz factor of the particle and
$p=2\pi \gamma a/\lambda$ is the undulator parameter, $a$ is the
amplitude of bending. 
If $\om_{0}\ll\Om_{\rm ch}$, then the frequencies of channeling and 
undulator radiation are well separated. 
In this case the characteristics of undulator radiation
are practically  independent on channeling
oscillations\cite{KSG1998,KSG1999}, and the operational principle of a
crystalline undulator is the same as for a conventional
one\cite{Alferov1989,Barbini1990,RullhusenArtruDhez}, in which 
the monochromaticity of the radiation is the result of a constructive 
interference of the photons emitted from similar parts of trajectory. 

It was shown in \cite{KSG1998,KSG1999} that a crystalline undulator
can operate provided the following conditions are met:
\begin{eqnarray}
\fl
\cases{
C=4 \pi^{2} \E a/\dUmax \lambda^{2} < 1 & \mbox {- stable channeling},
\\
d<a \ll \lambda  &\mbox {- large-amplitude regime}, 
\\
N = L/\lambda \gg 1 & \mbox {- large number of periods}, 
\\
L \sim \min\Bigl[\Ld(C),L_{\rm a}(\omega)\Bigr] 
&   \mbox {- account for dechanneling and photon attenuation}, 
\\
\Delta \varepsilon/ \varepsilon\ll 1 & \mbox {- low radiative losses}. 
\label{equation.1}
}
\end{eqnarray}
Earlier, it was demonstrated that these conditions can be achieved in
a positron-based undulator\cite{KSG1998,KSG1999,KSG2004_review}. 
In the present work, we demonstrate this for an electron-based one.
Prior to doing this let us present a short description of the physics 
lying behind these  conditions.

A stable channeling of a projectile in a periodically bent channel occurs 
if the maximal centrifugal force in the channel $F_{\rm cf}$ is less 
than the  maximal interplanar force $\dUmax$, 
i.e. $C\equiv F_{\rm cf}/\dUmax < 1$.
For an ultra-relativistic particle $F_{\rm cf}\approx \E/R_{\min}$, 
where $R_{\min}=\lambda^{2}/4\pi^{2}a$ is the minimum curvature radius 
of the channel with the profile of the periodical bendings given by 
$y(z)= a\sin (2 \pi z /\lambda)$ (see \fref{figure.1}).

The operation of a crystalline undulator must be considered in the
large-amplitude regime. 
Omitting the discussion (see \cite{KSG1998,KSG1999,KSG2004_review}) we note, 
that the limit $a/d>1$ accompanied by the condition $C\ll 1$ is mostly 
advantageous,  since in this case the typical frequencies of undulator 
and channeling  radiation are well separated, and the latter does not 
affect the parameters  of the former, whereas the intensity of the 
undulator radiation becomes comparable or higher than that of the 
channeling radiation. 
On the other hand, the inequality $a\ll \lambda$ means that the crystal 
structure is not destroyed and the deformation is an elastic one. 
Also, this results in moderate values of the undulator parameter, $p\sim 1$,
which ensure that the emitted radiation is of the undulator type
rather than of the synchrotron one. 
Consequently, the quantities $a$, $d$ and $\lambda$ must satisfy 
the double inequality presented in the second line in (\ref{equation.1}).

The term 'undulator' signifies that the number of periods is large,
$N\gg 1$. 
This inequality leads to a very peculiar pattern of a
spectral-angular distribution of the emitted radiation. 
Namely, for each value of the emission angle $\theta$ 
(measured with respect to the undulator axis) the spectrum consists 
of a set of narrow, well-separated peaks (harmonics). 
In a conventional undulator, based on the action of a magnetic (or electric) 
field \cite{Alferov1989,Barbini1990,RullhusenArtruDhez}, 
the beams of charged particles and photons move in vacuum. 
In such 'ideal' conditions the peak intensity of the emitted radiation is
proportional to $N^{2}$.

In a crystalline undulator the beams move in a medium and 
are affected by the dechanneling and photon attenuation effects,  
which noticeably reduce the $N^2$ increase of the peak intensity
\cite{KSG2001_Dech}. 
Due to collisions with the crystal constituents the channeling 
particle increases its transverse energy $\E_{\perp}$. 
At some point $\E_{\perp}$ exceeds the interplanar potential barrier 
and leaves the channel. 
The average interval for a particle to penetrate into a crystal 
until it dechannels is called the dechanneling length $\Ld$. 
In a straight crystal $\Ld$ depends on its type, the energy and
the type of a projectile.
In addition to these, in a periodically bent crystal $\Ld$ acquires 
the dependence on the parameter $C$\cite{KSG1999}.
The dechanneling length introduces a natural upper limit
on the length of a crystalline undulator: $L\le L_{\rm d}(C)$. 
The essential difference between an electron-based undulator and a
positron-based one lies in the fact that the dechanneling process for 
these projectiles occurs differently, resulting in a strong inequality 
$L^{\rm e^{-}}_{\rm d}\ll L^{\rm e^{+}}_{\rm d}$\cite{Kumakhov1986}. 
We discuss this important issue in more detail in
\sref{DechannelingLengths}. 

The photon attenuation stands for a decrease of the intensity of 
the emitted radiation in the crystal due to the processes of absorption 
and scattering. The interval within which
the intensity decreases by a factor of $e$ is called the attenuation
length, $L_{\rm a}(\om)$ \cite{ParticleDataGroup2006}. 
This quantity is tabulated for a number of elements and 
over wide ranges of photon frequencies \cite{Hubbel}. 
It is worth noting that for sufficiently high photon
energies ($\hbar\om\ge 10^{2}$ keV) the restriction on the length
of undulator due to the photon attenuation effect becomes much less
severe than to the dechanneling process \cite{KSG1998,KSG1999,KSG2004_review}. 

It was demonstrated \cite{KSG2001_Dech,KSG2005_SPIE1} that in the limit 
$L\gg \Ld(C)$  the intensity of radiation is defined not by the total number 
of undulator  periods $N=L/\lambda$, but rather by the quantity 
$\Nd=\Ld(C)/\lambda$ which is the number of periods within the 
dechanneling length.
Since for an ultra-relativistic particle $\Ld\propto \E$ 
\cite{Baier,BiryukovChesnokovKotovBook,Uggerhoj1980}, it seems natural 
that to increase $N_{\rm d}$ one can consider higher energies. 
However, at higher energies another limitation appears
\cite{KSG1998,KSG1999,KSG2000_loss}. 
The coherence of undulator radiation is only possible when the 
energy loss $\Delta \E$ of the particle during its passage through 
the undulator is small, $\Delta \E \ll \E$. 
This statement together with the fact, that for an ultra-relativistic 
projectile $\Delta \E$ is mainly due to radiation \cite{Baier}, 
leads to the conclusion that $L$ must be much
smaller than the radiation length $\Lrad$, which defines the mean
energy loss of an ultra-relativistic particle per unit length (see
e.g.\cite{Baier}). 
Therefore, we come to the forth and fifth limitations in (\ref{equation.1}).

A thorough analysis of the conditions (\ref{equation.1}) for the system
"periodically bent crystal + ultra-relativistic positrons" was
performed  for the first time in \cite{KSG1998,KSG1999} 
(and analyzed further in 
\cite{KKSG2000_tot,KSG2000_loss,KSG2004_review,KSG2005_SPIE1} ).
The ranges of $\varepsilon$, $a$ and $\lambda$ for a number of crystals were
established within which the operation of the crystalline undulator
is possible. 
These ranges include 
$\E=(0.5\ldots5)$ GeV, $a/d=10^{1}\dots10^2$, $C=0.01\dots 0.2$, 
$N\sim \Nd = 10^{1}\dots10^{2}$, $\omega\ge 10^{2}$ keV and are
common for all investigated crystals. 
The importance of the above mentioned regime in application to a 
positron-based undulator was later realized by other authors 
\cite{BellucciEtal2003,BaranovEtAl2005}.

In the case of electron channeling the restriction due to the dechanneling 
effect on the parameters of undulator are much more severe 
\cite{KSG1998,KSG1999,KSG2004_review}. 
Therefore it has been commonly acknowledged that the concept
of an electron-based undulator cannot be realized. 
In what follows we demonstrate that the crystalline undulator
of this type is feasible but operates 
in the regime of higher beam energies than the positron-based undulator.

\section{Dechanneling length of positrons and electrons in a periodically 
bent crystal \label{DechannelingLengths}}

As mentioned, random multiple scattering of a channeling particle by
electrons and nuclei of the crystal leads to the dechanneling effect.
As a result, the volume density $n(z)$ of the channeling  particles 
decreases with the penetration distance $z$. 
The dependence $n(z)$ can be approximated as $n(z)=n_{0}\exp(-z/\Ld)$ 
\cite{BiryukovChesnokovKotovBook,Kumakhov2,KSG1999}, 
where $n_{0}$ is the density at the entrance point.
It follows from here that the dechanneling length $\Ld(C)$ 
is a natural limitation for the effective length of a crystalline 
undulator \cite{KSG1999,KSG2001_Dech}. 
Therefore, the crystalline undulator is feasible when $L_{\rm d}(C)$ is 
large enough to ensure that the number of undulator periods 
within the dechanneling length satisfies the condition 
$N_{\rm d}=\Ld(C)/\lambda \gg 1$.

It is known that the dechanneling length of positrons exceeds that
of electrons (see, e.g., \cite{Kumakhov1986}). 
This is due to the difference in the channeling processes for 
these projectiles. 
Indeed, positrons, possessing positive charge, are repulsed by crystal 
atoms and, thus, channel between neighbouring crystallographic planes 
(see \fref{figure.1}(a)), where the concentration of nuclei and
electrons is low. 
In contrast, electrons channel in close vicinity of atomic planes 
(\fref{figure.1}(b)) where the number of the collisions with the 
crystal constituents is much larger. 
Therefore, electrons dechannel faster.

Let us discuss the approximations which one can use to calculate the
dechanneling lengths of a positron and an electron.

The influence of the dechanneling process on the photon
emission in a positron-based crystalline undulator  was considered in 
\cite{KSG1998,KSG1999,KSG2001_Dech}.
In the cited works the expression for the dechanneling length of 
ultra-relativistic positrons in a periodically bent crystal,
based of the diffusion model\cite{BiryukovChesnokovKotovBook}, 
was written as follows:
\begin{equation}
\Ld(C)=(1-C)^{2} \Ld(0), 
\label{equation.2}
\end{equation}
where $\Ld(0)$ is the dechanneling length a straight channel.
This quantity can be estimated as
$\Ld(0)=(256/9\pi^2)(a_{\rm TF}\, d \gamma/r_0\,\Lambda_{\rm c})$ 
\cite{KSG1999,BiryukovChesnokovKotovBook}, where $r_0$ and $a_{\rm TF}$ are 
the classical radius of the electron and the Thomas-Fermi radius of the
crystal atom, and $\Lambda_{\rm c}=\ln(\sqrt{2 \gamma} \, mc^{2}/I)-23/24$
($I$ stands for an average ionization potential of the atom).

The factor $(1-C)^{2}$ takes into account the decrease of the
interplanar potential well due to periodic bendings. 
The advantage of the approximate formula \eref{equation.2} is that 
it explicitly demonstrates the dependence of the dechanneling length 
on $C$, $\E$ and parameters of the channel. 
Equation (\ref{equation.2}) was tested in \cite{KSG2001_Dech} against more
 rigorous calculations, based on the simulation procedure of the positron 
channeling in straight and bent crystals. 
It was demonstrated that in wide range of the parameters \eref{equation.2} 
produces a good quantitative estimate for $L_{\rm d}(C)$.

To estimate the dechanneling length for electrons, one can use the
model due to Baier and Katkov \cite{Baier} which relates $\Ld(0)$ to 
a mean square of the multiple scattering angle of an 
ultra-relativistic electron. 
Since electrons channel in close vicinity of atomic planes, the 
multiple scattering occur predominantly from the nuclei of the
crystal. 
The latter makes the main contribution to the increase of
$\E_{\perp}$ \cite{Baier}. 
Let $q$ denote the mean square of the multiple scattering angle per 
a unit length. 
Then, $\Ld(0)$ can be defined as the length within which the mean 
square of the multiple scattering angle becomes equal to the square 
of Lindhard angle $\theta_{\rm L}$, i.e.
\begin{equation}
\Ld(0)=\theta_{\rm L}^{2}/q. 
\label{equation.3}
\end{equation}
Taking into account that individual scattering events are
independent and using the small-angle scattering approximation at
high energies\cite{Baier1}, one estimates $q$ as follows\cite{Baier}
\begin{equation}
q \simeq 
\frac{2 \pi m^{2} c^{4}}{\alpha \, \E^2} \,
L_{\rm r}^{-1}. 
\label{equation.4}
\end{equation}
Here $\alpha$ is the fine structure constant, $L_{\rm r}$ is the
radiation length in an amorphous medium. 
Using \eref{equation.4} in \eref{equation.3} and recalling that 
$\theta_{\rm L}=\sqrt{2 \Delta U/\E}$ ($\Delta U$ stands for the 
depth of the interplanar potential well),
one derives the following expression for the dechanneling length of
an ultra-relativistic electron in a straight crystal \cite{Baier}
\begin{equation}
\Ld(0)\simeq  {\alpha \over \pi} {\Delta U\,\E \over  m^2 c^4} \,
\Lrad.
\label{equation.5}
\end{equation}
This equation has been derived assuming that the
dechanneling length of ultra-relativistic electrons in a straight
crystal is less than the characteristic length of the radiation
losses \cite{Baier}.

As far as we know (see also \cite{Carrigan2006}),
experimental measurements of the dechanneling length during planar 
channeling of high energy electrons were performed in 
\cite{AdejshviliEtAl1984,AdejshviliEtAl1985,KomakiEtAl1984}. 
In \cite{AdejshviliEtAl1984,AdejshviliEtAl1985} the dechanneling length 
for 1.2 GeV electrons, channeled in  Si(110) was determined as 
$\Ld(0)=25\pm5 \, \mu$m. 
This value, as indicated in \cite{AdejshviliEtAl1984}, is in a good
agreement with  \eref{equation.5}. 
However, in the experiment \cite{KomakiEtAl1984} the dechanneling length 
for 350 MeV electrons channeled in Si(110) was found as 
$\Ld(0)=31 \, \mu$m, which is even greater than the value based on 
(\ref{equation.5}). 
Therefore, we believe that the approximate formula \eref{equation.5} can 
be used for a quantitative estimate of the dechanneling length of
ultra-relativistic electrons in straight crystals.

\begin{figure}[ht]
\centering
\includegraphics[clip,scale=0.5,angle=0]{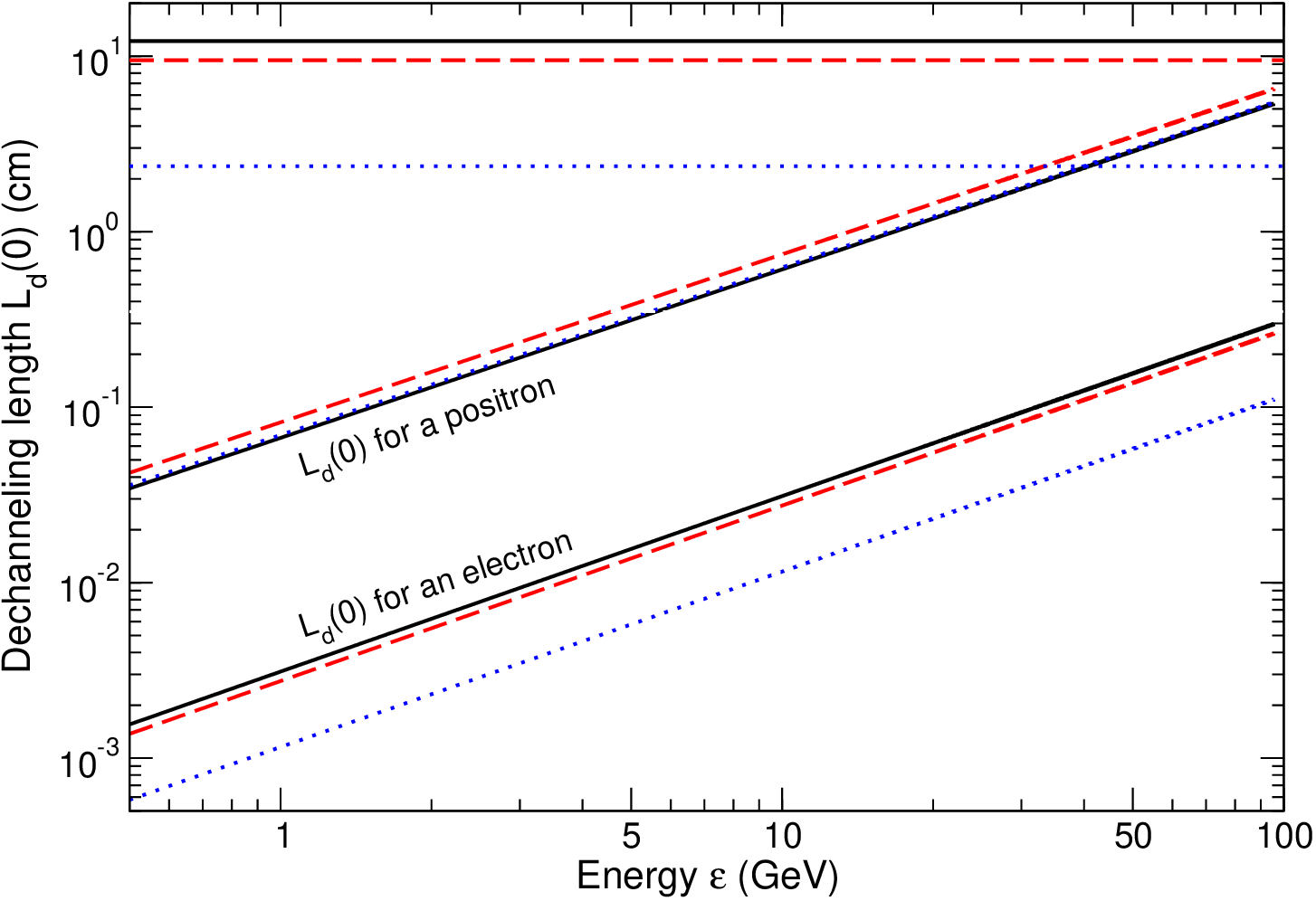}
\caption{
Dependence of the dechanneling length of
ultra-relativistic electrons and positrons on energy in 
straight (111) planar channels of 
C (solid line), Si (dashed line), Ge(dotted line).
Horizontal lines denote the radiation lengths in the corresponding crystals.
}
\label{figure.2}
\end{figure}

\Fref{figure.2} presents the dependence of $\Ld(0)$ on $\E$ 
for planar channeling of electrons, calculated in accordance 
with \eref{equation.5}, and of positrons (see \eref{equation.2} for $C=0$) 
in various straight  channels.
Horizontal lines show the radiation length in the crystals. 
To calculate $L_{\rm r}$ we used equation (27.22) from 
\cite{ParticleDataGroup2006}. 
It is seen from \fref{figure.2} that for all energies the dechanneling
length for a positron exceeds that for an electron by more than an
order of magnitude: 
$L_{\rm d}^{\rm e^{+}}/L_{\rm d}^{\rm e^{-}}\sim 10^1\dots 10^2$. 
A positron-based crystalline undulator can be considered for 
$\E\le 10$ GeV, where the radiation length greatly exceeds  $\Ld$. 
One can demonstrate that in this case it is possible to achieve 
$\Nd \sim 10^1\dots 10^2$ \cite{KSG2004_review,KSG2005_SPIE1}. 
The corresponding values of the undulator period are 
$\lambda=\Ld/\Nd = 10^{-4}\dots 10^{-2}$ cm. 
From \fref{figure.2} it is seen that this is exactly the interval to 
which $L_{\rm d}^{\rm e^{-}}$ belongs. 
Therefore, for an electron of the energy much smaller than
10 GeV the number of undulator periods within the dechanneling length is
equal, in the order of magnitude, to one. 
Hence, an electron-based crystalline undulator can hardly be realized 
in this energy regime.

For higher energies, $\E > 10$ GeV, the dechanneling length
of a positron becomes comparable with the radiation length. 
This means, that a positron-based crystalline undulator cannot be
realized due to large radiation losses (see the last condition in
\eref{equation.1}). 
On the other hand, the dechanneling length of electrons at
such energies, being much lower than $L_{\rm r}$,
becomes comparable with $\Ld$ for positrons but of lower
energies. 
Therefore, an electron-based crystalline undulator is
meaningful to discuss within the interval 
$\E\sim 10^1\dots 10^2$ GeV.

To conclude this section let us discuss a model which allows one to
define the dechanneling length of an electron in a {\it periodically
bent} crystal.

The factor $(1-C)^{2}$ in \eref{equation.2} appears as a result of application 
of the harmonic approximation for interplanar potential
$U(x)$ \cite{KSG2000_loss,KSG2004_review,BiryukovChesnokovKotovBook}. 
This approximation is adequate for a positively charged projectile, 
but its validity is not obvious for an electron. 
In the latter case the interplanar potential is
strongly anharmonic \cite{Kumakhov1986}. 
To calculate the dependence of dechanneling length on $C$ one can 
consider the following arguments. 
In the point of maximum curvature, the effective potential,
acting on the electron, can be written as 
$U_C(x)=U(x)-C \dUmax\, x$, where $x$ is the distance from the plane.
The depth of the effective potential well, $\Delta U_C$, 
defines the maximum value of the transverse energy which an electron 
may gain. 
Within the framework of the diffusion theory \cite{BiryukovChesnokovKotovBook} 
the dechanneling length $\Ld(C)$ of an ultra-relativistic projectile 
in a bent channel is proportional to $\Delta U_{\rm C}$. 
Hence one can write $\Ld(C)=k(C)\, \Ld(0)$, 
where $\Ld(0)$ is defined by equation (\ref{equation.5}) and $k(C)$ stands
for the ratio $\Delta U_{\rm C}/\Delta U<1$.
To obtain the explicit dependence $k(C)$ one has to calculate the 
quantity $\Delta U_{\rm C}$ using a realistic model for $U(x)$.

\begin{figure}[ht]
\centering
\includegraphics[clip,scale=0.5,angle=0]{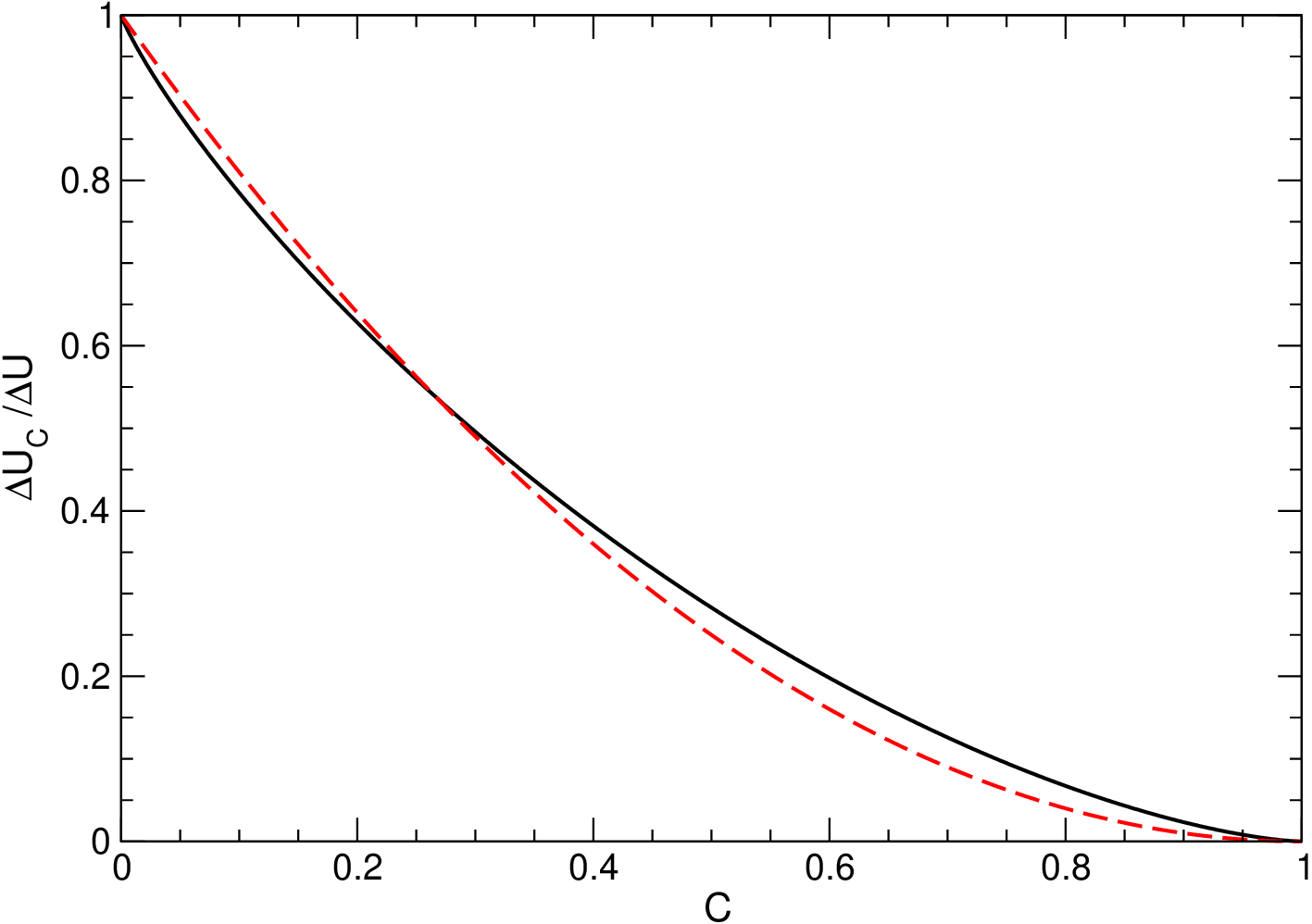}
\caption{$\Delta U_{C}/\Delta U$ versus $C$ for the P\"oschl-Teller
potential (solid curve) and $(1-C)^{2}$ (dashed curve).}
\label{figure.3}
\end{figure}

In our estimates we used the P\"oschl-Teller potential \cite{Baier}. 
The results of calculation are presented in \fref{figure.3}, where the
solid curve stands for $k(C)$ within the  P\"oschl-Teller model.
The dashed curve presents the dependence $(1-C)^{2}$. 
It is seen that both curves are close for all $C$. 
Therefore, to estimate the dechanneling length of an ultra-relativistic 
electron in a periodically bent crystal one can use equation
\eref{equation.2} with $\Ld(0)$ from \eref{equation.5}. 
Consequently, the condition of a large number of periods within 
the dechanneling acquires the form
\begin{equation}
\Nd = (1-C)^{2} \Ld(0)/\lambda\gg 1 \, .
\label{equation.6}
\end{equation}

In the next section we use equations \eref{equation.2}, \eref{equation.5} 
and \eref{equation.6} to calculate the characteristics of an electron-based 
crystalline undulator.

\section{Numerical results for the electron-based crystalline undulator
\label{NumericalResults}}

To prove the feasibility of an electron-based crystalline undulator
for a fixed value of the undulator periods $\Nd = \Ld(C)/\lambda$ 
it is necessary to establish the ranges of $\E$, $a$ and $\lambda$ 
within which all the conditions, formulated in \eref{equation.1}, are met. 
Once the ranges are found one can calculate the spectral-angular 
distribution of the energy emitted in the undulator. 
The results, presented below in this section, show, that for an 
electron-based crystalline undulator the energy of emitted photons 
is in the range $\hbar \om\ge 10^{2}$ keV. 
For such energies the values of the attenuation length for all 
crystals fall within the $cm$ range \cite{Hubbel}, resulting in a strong 
inequality $\La(\om) \gg \Ld(0)$. 
Therefore, it is the dechanneling which becomes the dominant effect 
restricting the length of the undulator (see the forth condition in 
\eref{equation.1}).

To perform the numerical analysis let us express $C$, $a$ and
$\om$ as functions of $\lambda$. 
From  $\Nd=(1-C)^{2} \Ld(0)/\lambda$ one finds the following 
expression for the dependence $C(\lambda)$:
\begin{equation}
C(\lambda) 
= 
1 - \sqrt{{\Nd \lambda \over \Ld(0)}}.
\label{equation.7}
\end{equation}
Because of the condition  $C\ge 0$, the quantity 
$\lambda_{\max}=\Ld(0)/\Nd$ defines the maximum value of the undulator 
period.

Substituting $C=4 \pi^{2} \E a/\dUmax\lambda^{2}$
into (\ref{equation.7}) one derives the dependence $a(\lambda)$:
\begin{equation}
a(\lambda) 
= 
{\lambda^{2} \dUmax \over 4 \pi^{2} \E} \,
\left(1-\sqrt{{\Nd \lambda \over \Ld(0)}}
\right) \,.
\label{equation.8}
\end{equation}
From \eref{equation.7} and \eref{equation.8} follows that by tuning 
$\lambda$ and $\E$ for fixed $\Nd \gg 1$, one can establish the
ranges of $a$ and $C$ where the first and second conditions in
\eref{equation.1} are met for a given crystal.

Using the dependence $a(\lambda)$ it is possible to calculate the
frequency of the fundamental harmonic as a function of $\lambda$:
$\om_{1}(\lambda)=8 \pi \gamma^{2} c\lambda^{-1}/(2+p^{2}(\lambda))$ 
(here $p(\lambda)=2 \pi \gamma a(\lambda)/\lambda$ is the undulator parameter). 
The spectral-angular distribution of undulator radiation in the 
forward direction at $\omega=\omega_{1}$ is calculated as follows 
\cite{Alferov1989,Baier,KSG2005_SPIE1}:
\begin{equation}
\left. 
{{\rm d^{3}} E \over \hbar {\rm d} \om {\rm d} \Om} 
\right|_{\om=\om_{1} \atop \theta=0^{\circ}}
=
4 \alpha
\gamma^{2} N_{\rm d}^{2} z (1-2z)
\Bigl[J_{0}(z)-J_{1}(z)\Bigr]^{2}\,, 
\label{equation.9}
\end{equation}
where $z=p^{2}/2(2+p^{2})$, $J_{n}(z)$ is the Bessel function 
of integer order and $\theta$ is the emission angle with respect 
to the undulator axis.

\begin{figure}[ht]
\centering
\includegraphics[clip,scale=0.5,angle=0]{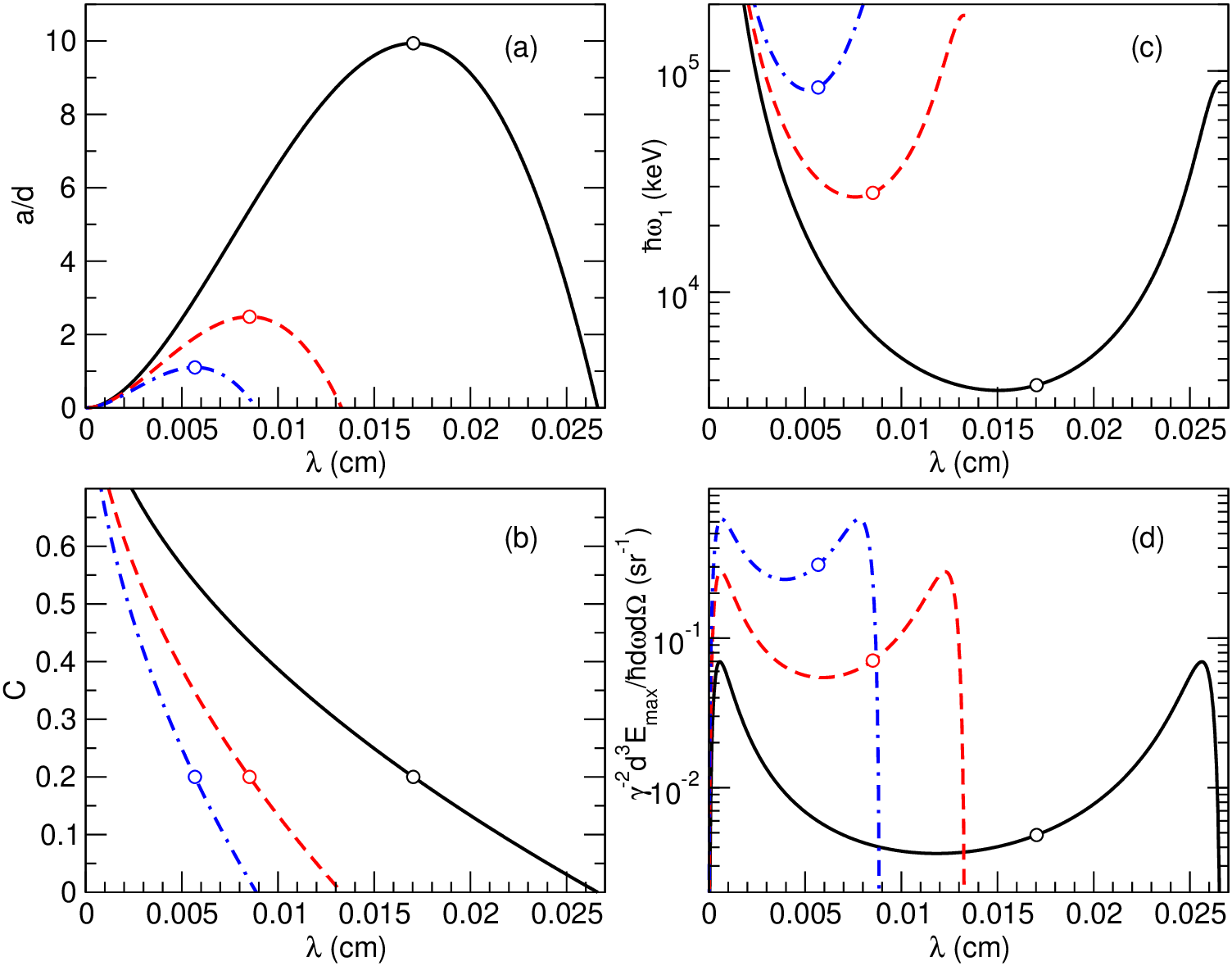} 
\caption{
Dependences of $a/d$, $C$, $\hbar\omega_{1}$ and the peak intensity 
(\ref{equation.9}) on $\lambda$ for a $50$ GeV electron channeling in 
a periodically bent Si(111) ($d=2.35$ \AA). 
In each graph the three curves correspond to different
values of undulator periods within the dechanneling length: 
the solid curves stand for $N_{\rm d}=5$, 
the dashed curves - for $N_{\rm d}=10$, 
the chained curves - for $N_{\rm d}=15$.
For each $N_{\rm d}$ the open circles indicate the parameters of
undulator with $C=0.2$ (see graph (b)).
This $C$ value ensures the maximum of the ratio $a/d$
(graph (a)). 
}
\label{figure.4}
\end{figure}

\begin{figure}[ht]
\centering
\includegraphics[clip,scale=0.5,angle=0]{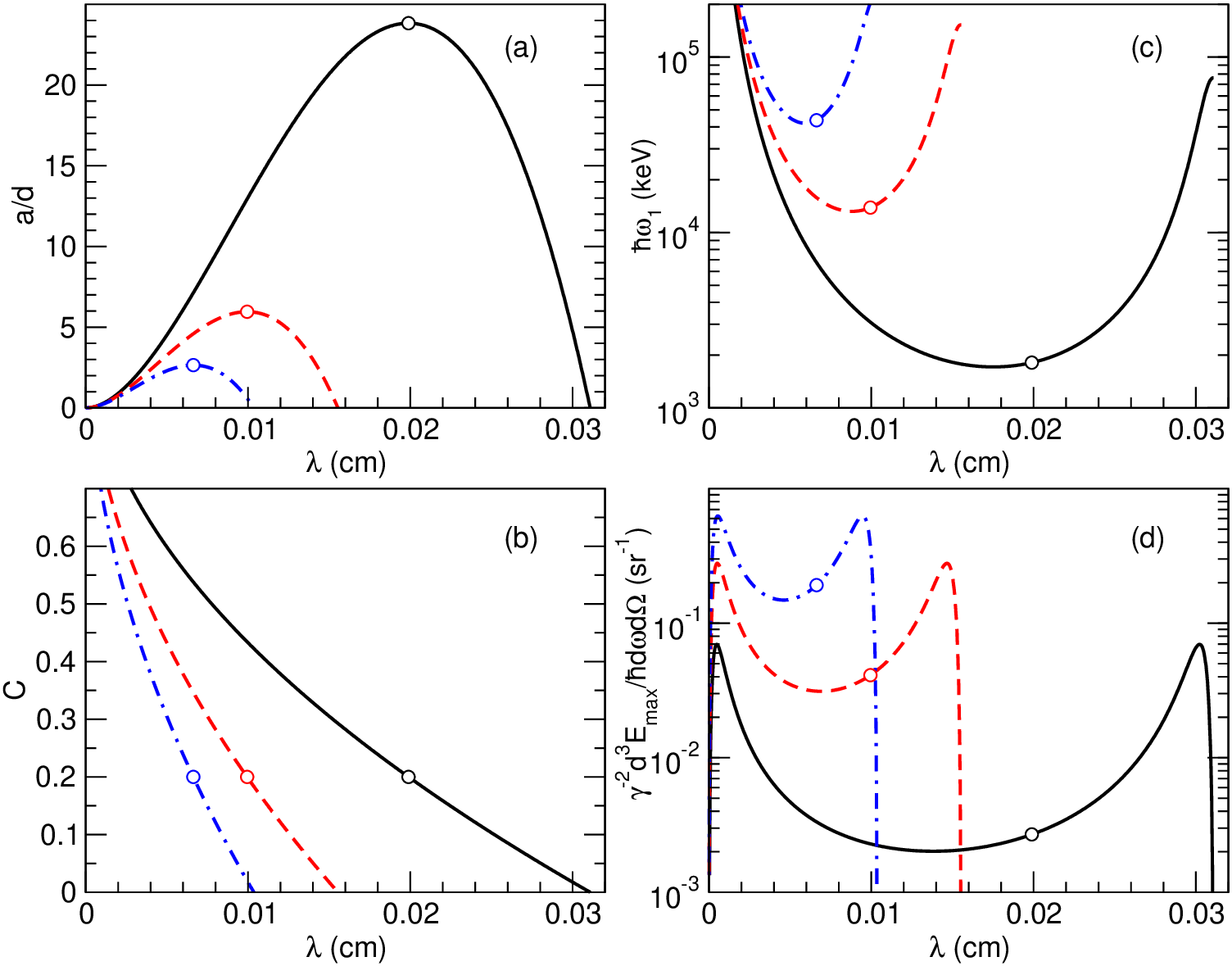}
\caption{Same as in figure \ref{figure.4} but for C(111).} 
\label{figure.5}
\end{figure}

Figures \ref{figure.4} and \ref{figure.5} present the results of
numerical calculations of the dependences $a$, $C$, $\hbar\om_1$
and ${\rm d}^{3}E/\gamma^{2} {\rm d}\hbar\omega {\rm d}\Om$ on
$\lambda$ for $\E=50$ GeV electrons channeling in Si(111)
and C(111). 
In figures \ref{figure.4}(a) and \ref{figure.5}(a) 
the ratio $a/d$ versus $\lambda$ is shown for the fixed values of undulator
periods within the dechanneling length. These graphs illustrate the
ranges of $a$, $\lambda$ and $N_{\rm d}$ within which the second and
third conditions from \eref{equation.1} are met. The curves presented in
the figures suggest that the condition $a/d>1$ is fulfilled for
$N_{\rm d}\le 15$, i.e. the undulator with a sufficiently large
number of periods can be considered.

%

Comparing figures \ref{figure.4}(a) and \ref{figure.5}(a) 
one notices that the curves for C(111) produce higher values of 
the ratio $a/d$ than those for Si(111) calculated for the same 
number of periods. 
Let us explain this difference. 
It follows from \eref{equation.8} that the ratio $a/d$ as a function of 
$\lambda$ attains maximum at $\lambda=16\lambda_{\rm max}/25$. 
The maximum value of the ratio is given by
\begin{equation}
\left({a \over d} \right)_{\rm max} 
=
{4^3 \over 5^5 \pi^2 \Nd^2 \, \E}
{\dUmax L_{\rm d}^2(0) \over d}.
\label{equation.10}
\end{equation}
For fixed values of $\varepsilon$ and $N_{\rm d}$ the magnitude of
$(a/d)_{\rm max}$ depends on the parameters of a channel, $\dUmax$ and $d$, 
and on the dechanneling length $\Ld(0)$, equation \ref{equation.2}.
Taking into account that  $\dUmax$ and $d$ equal to $9.2$~GeV/cm and 
$1.54$~\AA\, for C(111) and to $8.0$ GeV/cm and $2.35$~\AA\,  for Si(111) 
(see \cite{BiryukovChesnokovKotovBook}), and $\Ld(0)\approx 0.16$ cm for C and
$\Ld(0)\approx 0.13$ cm for Si (see figure \ref{figure.2}), 
one finds that the ratio
$(a/d)_{\rm max}$ for C(111) is approximately $2.4$ times higher
than that for Si(111).

One can easily demonstrate that the ratio $(a/d)_{\max}$ is reached 
when $C=0.2$. 
This is valid for all $N_{\rm d}$ (see figures \ref{figure.4} and 
\ref{figure.5}, where open circles mark the parameters corresponding 
to this value of $C$).
The graphs (a) and (b) in figures \ref{figure.4} and \ref{figure.5} 
allow one to explicitly establish the ranges of parameters within which 
the conditions \eref{equation.1} are fulfilled,
and consequently, the operation of an electron-based undulator is feasible. 
These ranges are: $C\le 0.2$, $\Nd \simeq 10$,
$\lambda\approx 10^1...10^2\, {\rm \mu m}$ and $a\approx 2...20$ {\AA}. 
Let us note that the indicated ranges of $\lambda$ and $a$ are close to those 
which were established for a positron-based undulator (see, e.g., \cite{KSG1999}).

Graphs (c,d) in figures \ref{figure.4} and \ref{figure.5} 
present the dependences on $\lambda$ of the energy fundamental harmonic 
(graph (c)) and of the peak intensity \eref{equation.9} calculated 
in the forward direction at $\om=\om_1$ and scaled by the factor 
$\gamma^{2}$, graph (d).
These graphs demonstrate that within the $a$, $\lambda$ and $\Nd$ ranges 
indicated above, the magnitude of $\hbar\om$ and of
the intensity of undulator radiation can be varied by the orders of
magnitude.

Figures \ref{figure.4}(c) and \ref{figure.5}(c) 
indicate that the energy of photons emitted in the $50$ GeV 
electron-based crystalline undulator lies within the $1 \dots 10^{2}$ 
MeV range. 
Let us note here, that the corresponding values of the attenuation lengths 
(for both C and Si crystals) exceed $5$ cm (see, e.g. 
\cite{ParticleDataGroup2006,Hubbel}). 
This value by far exceeds the dechanneling lengths in the crystals.
Therefore, the statement on the negligible role of the photon attenuation, 
made in the opening paragraph of this section, is fully justified.

\begin{figure}[ht]
\centering
\includegraphics[clip,scale=0.5,angle=0]{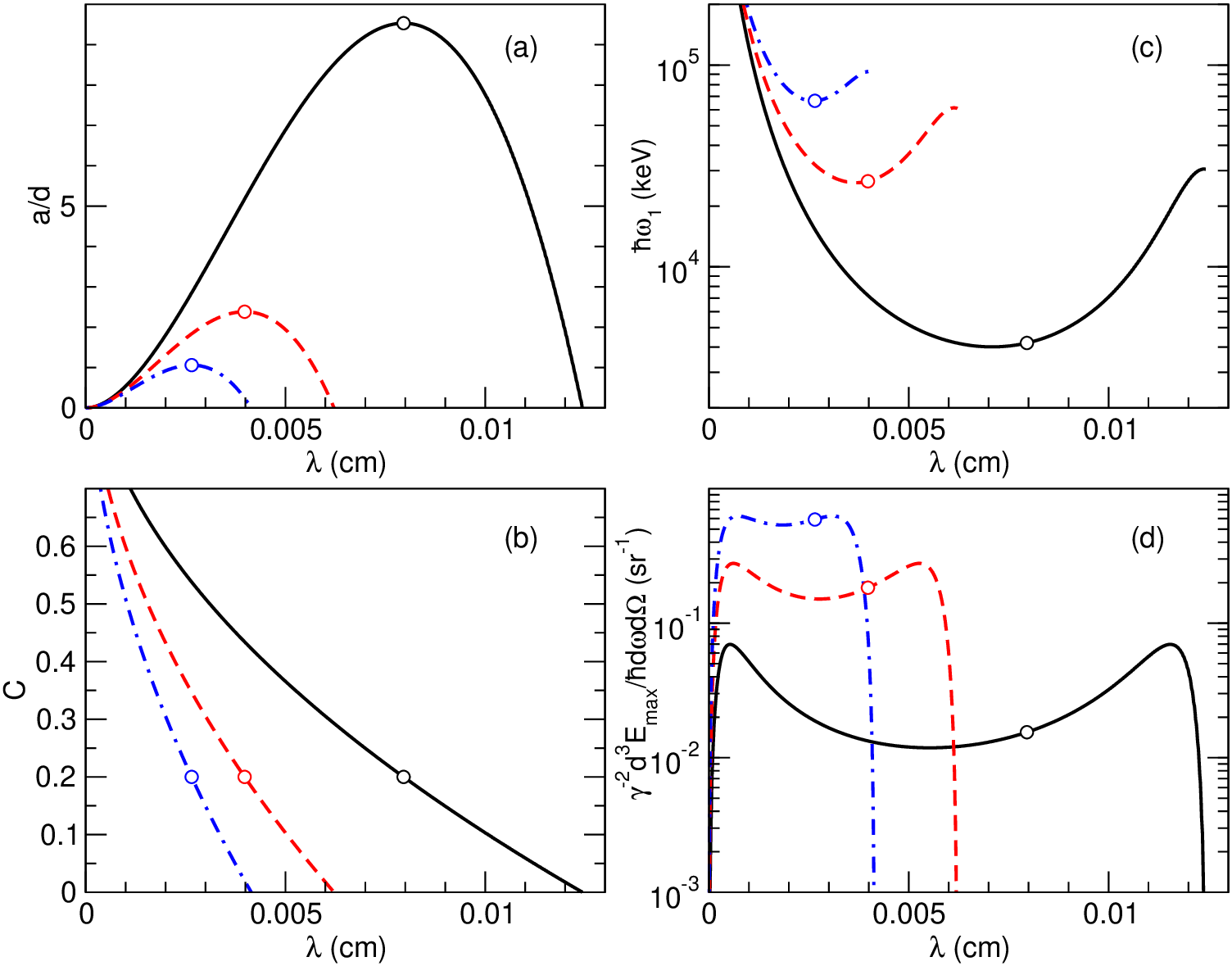}
\caption{Same as in figure \ref{figure.4} but for a 20 GeV
electron channeling in the periodically bent C(111).} 
\label{figure.6}
\end{figure}

Let us now discuss the extent to which the decrease of the electron energy
influences the allowed ranges of parameters of the crystalline undulator. 
\Fref{figure.6} presents the dependences of $a/d$, $C$,
$\hbar\om_1$ and the peak intensity on $\lambda$ for the electron
of a lower energy, $\E=20$ GeV, channeling in C(111).

Comparing the curves in figures \ref{figure.6}(a,b) with the
corresponding dependences from figures \ref{figure.5}(a,b) one notices, that 
the domain of parameters $a/d$, $\lambda$ and $\Nd$ consistent with the 
conditions from \eref{equation.1} shrinks with the decrease of $\E$. 
Firstly, it is seen that the undulator period for a $20$ GeV electron is 
noticeable smaller than that for a $50$ GeV one. 
This feature is a corollary of a linear dependence of the dechanneling length 
on $\E$ (see \eref{equation.5}). 
As a result, the maximum value of the undulator period,
 $\lambda_{\max}=\Ld(0)/\Nd$, consistent with the condition $0 \le C\le 1$, 
is $2.5$ times less for a $20$ GeV electron. 
Due to the same reason the maximal values of $a/d$ in \fref{figure.6}(a) 
are $2.5$ times lower than those in \fref{figure.5}(a). 
Indeed from \eref{equation.5} and \eref{equation.10} follows, 
that for fixed $\Nd$ and for the same channel 
$(a/d)_{\max}\propto L_{\rm d}^{2}(0)/\E\propto\E$. 
As a result, in the case of $20$ GeV electrons the condition $a/d>1$ 
can be satisfied only within the reduced interval of the undulator
periods, $\Nd<10$.

\begin{figure}[ht]
\centering
\includegraphics[clip,scale=0.4,angle=0]{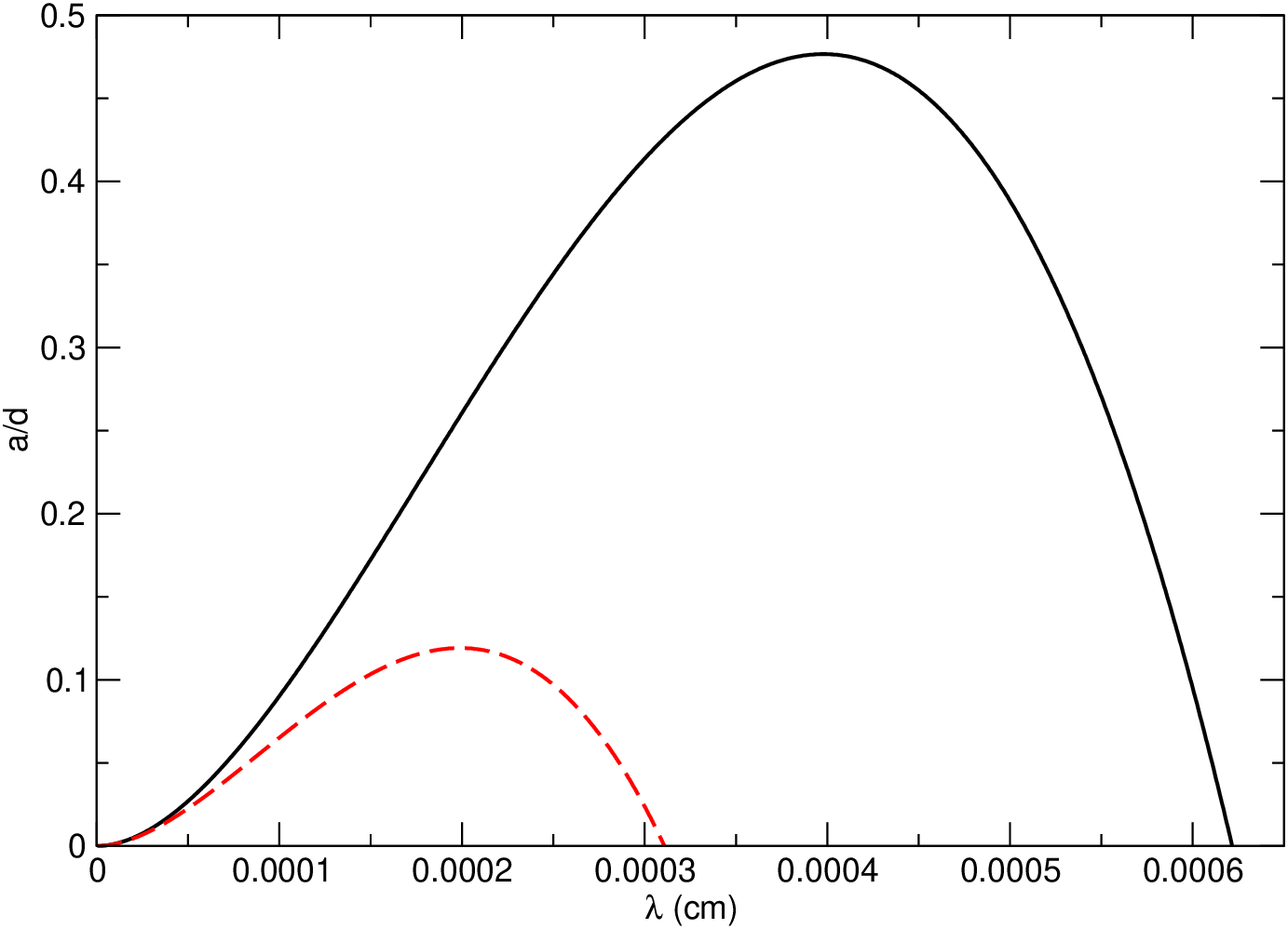}
\caption{Dependence of $a/d$ on $\lambda$ for a 1 GeV electron
channeling in a periodically bent C(111). 
The curves correspond to different values of undulator 
periods within the dechanneling length: the solid curve stands 
for $\Nd=5$, the dashed curve - for $\Nd=10$.
}
\label{figure.7}
\end{figure}
The arguments, presented above, indicate that further decrease of
$\E$ will result, eventually, in a collapse of the domain of
the parameters consistent with \eref{equation.1}. 
To illustrate this, in \fref{figure.7} we present the dependence of the 
ratio $a/d$  on $\lambda$ for a $1$ GeV electron. 
This figure illustrates that the case $\Nd\gg 1$ can be realized only 
if $a/d<1$, which contradicts to the second condition from \eref{equation.1}. 
For such low electron energies the large-amplitude regime can be realized only
for $\Nd\sim 1$.

\begin{figure}[ht]
\centering
\includegraphics[clip,scale=0.5,angle=0]{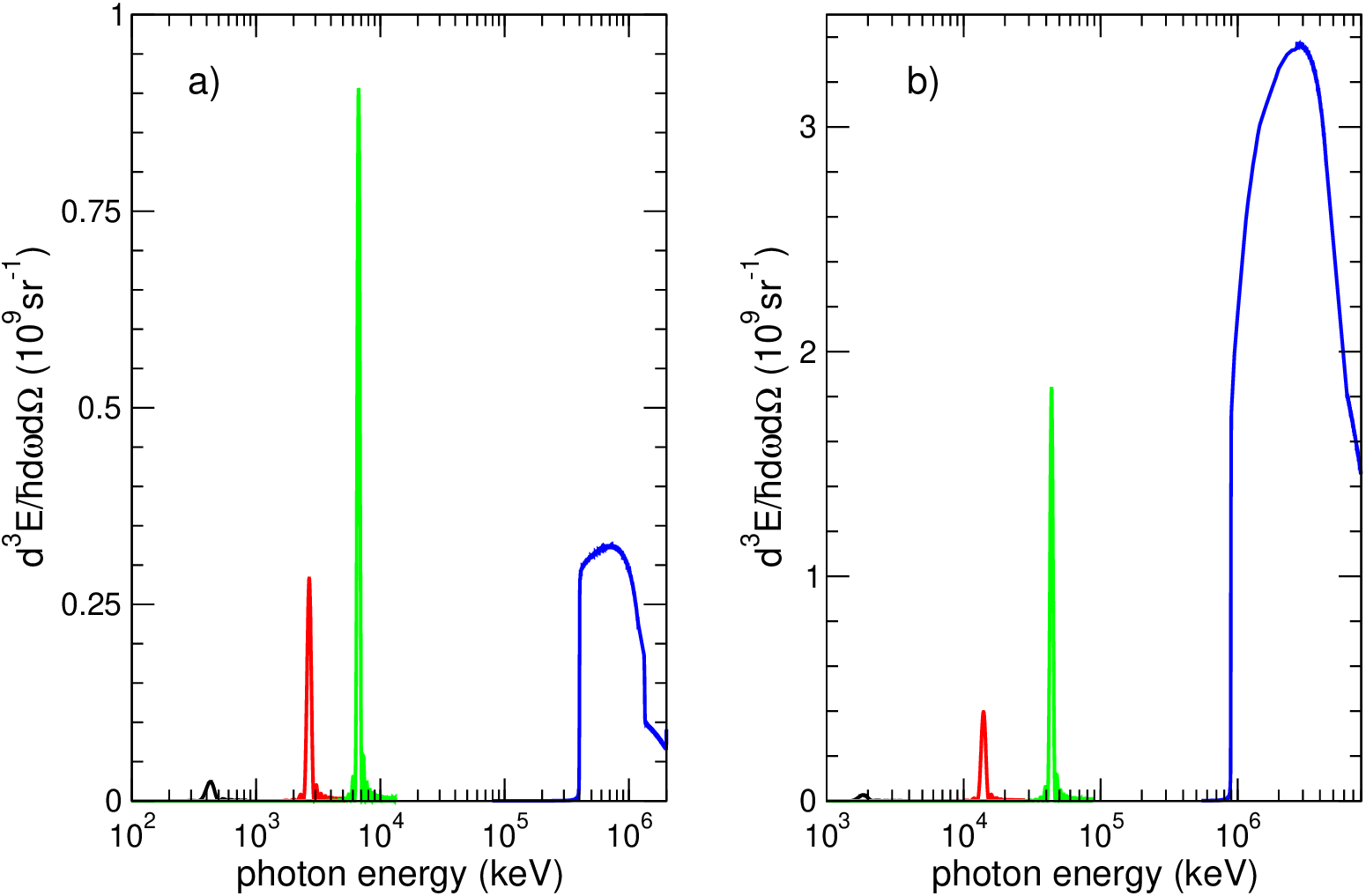}
\caption
{Spectral distributions of the undulator and channeling radiation 
emitted in the forward direction by a $20$ GeV (figure (a)) and 
by a $50$ GeV  (figure (b)) electron in C(111).
Narrow peaks stand for the spectral distribution of the
undulator radiation in the vicinity of the {\it fundamental} harmonics
for six different undulators (corresponding to $C=0.2$) 
defined by open circles in figures \ref{figure.5} and 
\ref{figure.6}.
In each graph the first (i.e., the leftest) narrow peak corresponds 
to $\Nd =5$,  the second peak - to $\Nd =10$, and the third peak 
- to $\Nd =15$.
}
\label{figure.8}
\end{figure}

Figures \ref{figure.4}-\ref{figure.6} allow one to define a set 
of parameters which characterize the undulator and its radiation.
For example, fixing $N_{\rm d}$ and $C$ one finds: 
the period $\lambda$ \-- from graphs (b),
the amplitude $a$ \-- from graphs (a),
$\hbar\om_1$  and the peak intensity 
\--  from graphs (c) and (d).
In each figure, open circles mark the parameters of undulators with  
different number of $N_{\rm d}$ but with the same value of $C$
equal to 0.2.
For the undulators, based on the electron channeling in C(111),
we calculated the spectral distribution of radiation 
(in the forward direction) 
in vicinity of the corresponding {\it fundamental} harmonics, 
i.e. for $\om \sim \om_1$.
To calculate the distributions we followed the formalism, developed in
\cite{KSG2005_SPIE1}, which describes the undulator radiation in presence of 
the dechanneling and the photon attenuation.  
Narrow peaks in  figure \ref{figure.8} represent the results of 
these calculations.
Wide peak in each graph stands for the  spectral distribution of the 
channeling radiation in the forward direction.
To obtain the latter we, at first, calculated the spectra for individual 
trajectories (using the P\"oschl-Teller model \cite{Baier} for the 
interplanar potential), corresponding to a stable channeling for 
given $C$.
Then, the averaging procedure was carried out to calculate the spectra
(see \cite{KSG2000_loss,KKSG2000_tot} for the details).
Figure \ref{figure.8} clearly demonstrates  
that by tuning the parameters of bending and varying the electron energy
it is possible to separate the frequencies of the undulator radiation 
from those of the 
channeling radiation, and to make the intensity of the former comparable 
or higher than of the latter. 

\section{Conclusion \label{Conclusion} }

We have demonstrated that it is feasible to devise an undulator 
based on the channeling effect of ultra-relativistic electrons in
a periodically bent crystal.
The electron-based undulator operates in the tens of GeV range of electron 
energies.
These energies are noticeably higher than those in a positron-based
undulator. 
Apart from the difference in energies of the projectiles, 
other parameters of the crystalline undulators (i.e., $a$, $p$, $\lambda$)
are much alike.
Therefore, to construct an electron-based undulator 
one can consider the methods proposed earlier in connection with a 
positron-based undulator.
These methods include propagation of an acoustic wave 
\cite{KSG1998,KSG1999,Kaplin} or the use of a graded composition of different 
layers \cite{MikkelsenUggerhoj2000} or a periodic mechanical deformation of 
the crystalline structure \cite{BellucciEtal2003}.

{Present technologies allow one to construct the  
periodically bent crystalline structures with the required parameters
\cite{PECU}.
Similar to the case of a positron-based undulator \cite{KSG2005_SPIE1,PECU}, 
the parameters of high-energy electrons beams available 
at present (see Ch. 26 in \cite{ParticleDataGroup2006})
are sufficient to achieve the necessary conditions to construct the 
undulator and to create, on its basis,
powerful radiation sources in the $\gamma$-region of the spectrum.
}
As in the positron case \cite{KSG1998,KSG1999,KSG2004_review} it is meaningful
to explore the idea of a $\gamma$-laser 
by means of an electron-based  undulator.

This work has been supported by the European Commission 
(The PECU project, contract No. 4916).

\end{document}